\documentclass[fleqn,twoside]{article}
\usepackage{fleqn,espcrc2,epsf}

\usepackage{graphicx}

\newcommand{\AmS}{{\protect\the\textfont2
  A\kern-.1667em\lower.5ex\hbox{M}\kern-.125emS}}

\newcommand{\nn}{\nonumber}
\def\cancel#1#2{\ooalign{$\hfil#1\mkern1mu/\hfil$\crcr$#1#2$}}
\def\slash#1{\mathpalette\cancel{#1}}

\title{Staggered Fermion Thermodynamics using Anisotropic Lattices
\thanks{This work was conducted on the QCDSP machines at Columbia
     University and the RIKEN-BNL Research Center, in collaboration
     with Thomas Manke and members of the RBC Collaboration. LL is
     supported by the US DOE.
}}

\author{L.~Levkova\address{Department of Physics, Columbia 
University, New York, NY, 10027}}
\begin{document}
\bibliographystyle{apsrev}

\begin{abstract}
Numerical simulations of full QCD on anisotropic lattices provide a 
convenient way to study QCD thermodynamics with fixed physics scales 
and reduced lattice spacing errors. We report results from calculations 
with 2-flavors of dynamical fermions where all bare parameters and hence 
the physics scales are kept constant while the temperature is
changed in small steps by varying only the number of the time slices. 
The results from a series of zero-temperature scale setting simulations 
are used to determine the Karsch coefficients and the equation
of state at finite temperatures.

\vspace{1pc}
\end{abstract}

\maketitle

\section{INTRODUCTION}
The anisotropic formulation of lattice QCD has certain advantages when it comes to
 the study of the equation of state (EOS) at various finite temperatures. 
The finite temperature field theory has a 
natural asymmetry which makes the anisotropic approach useful to reduce the 
lattice spacing errors associated with the transfer matrix at less cost than 
is required for the full continuum limit\cite{finT}.
Through the introduction of anisotropy on the lattice one can make the temporal lattice spacing,
 $a_t$, sufficiently small so that by varying only the number of time slices, $N_t$, the temperature 
can be changed in small discrete steps. 

Our simulations of full QCD with two flavors of staggered fermions on anisotropic 
lattices are aimed at the study of the thermodynamic properties of the quark-gluon system. 
We employ a fixed parameter
 scheme in which all the bare parameters of the simulation are kept constant and only the 
temperature is changed by varying $N_t$ (from 4 to 64). This approach separates temperature and lattice 
spacing effects and keeps the underlying physics scales fixed.

The calculation of the EOS of the quark-gluon system involves derivatives 
of the bare parameters with respect to the physical anisotropy $\xi = a_s/a_t$ and the spatial lattice spacing 
$a_s$. These
``Karsch'' coefficients\cite{karsch} are determined as a by-product of the zero-temperature
calculations needed to choose the bare parameters. Once determined, the Karsch coefficients
 can be used for all temperatures since they depend only on the intrinsic lattice
parameters and not on $N_t$.  This allows a straight-forward determination 
of the temperature dependence of the energy and pressure, again at fixed
lattice spacing. With two or more slightly different values for $a_t$, a high-resolution 
sampling of temperatures can be investigated.

\section{THE ANISOTROPIC STAGGERED ACTION AND EOS DERIVATION}
Our calculations are based on the QCD action
$S^{\xi} = S_G^{\xi} +  S_F^{\xi}$,  where the gauge action is:
\begin{eqnarray}
\label{eq:asym_gauge_action}
S_G^{\xi} 
  = -\frac{\beta}{N_c}\frac{1}{\xi_o}\left[
     \sum_{x, s>s^\prime} P_{ss^\prime}(x)  +
  \xi_o^2 \sum_{x, s} P_{st}(x) \right],
     \nonumber
\end{eqnarray}
and the fermion action is:
\begin{eqnarray} 
\label{eq:aniso_lattice_quark_action}
S_F^{\xi}
  &=& \sum_{x} \overline{\psi}(x) \left[
      m_f + 
      \nu_t  \slash{D}^{\rm Staggered}_t\right] \psi(x)\nonumber\\
& & +
     \sum_{x} \overline{\psi}(x)\left[  \frac{1}{\xi_0} \sum_s \slash{D}^{\rm Staggered}_s 
      \right]  \psi(x) . \nonumber
\end{eqnarray}

To derive the EOS we start from the thermodynamics identities
$\varepsilon(T)=-\frac{1}{V_s}\left.\frac{\partial\ln Z}{\partial (1/T)}\right|_{V_s}$ and $
p(T)=T\left.\frac{\partial\ln Z}{\partial V_s}\right|_{T}.$

Using the explicit form for $Z$ and changing the independent variables with respect to which we differentiate to $\xi$ and $a_s$ we get:
\footnotesize{
\begin{eqnarray}
\varepsilon(T)&=&\frac{\xi}{N_s^3N_ta_s^3a_t}\frac{1}{N_c}\times\nn\\
\hspace{-1cm}&&\left[\left(\frac{1}{\xi_o}\left.\frac{\partial \beta}{\partial \xi}\right|_{a_s}+\beta\left.\frac{\partial \xi_o^{-1}}
{\partial \xi}\right|_{a_s}\right)
\langle \sum_{x, s>s\prime} {\rm Re} {\rm Tr}\left[ P_{ss\prime}(x)\right]\rangle\right. \nonumber\\
&&\left.+\left(\xi_o\left.\frac{\partial \beta}{\partial \xi}\right|_{a_s}+\beta\left.\frac{\partial\xi_o}{\partial \xi}\right|_{a_s}\right)\langle\sum_{x, s} {\rm Re} {\rm Tr}\left[ P_{st}(x)\right]\rangle \right] \nonumber\\
&&+\frac{\xi N_f}{4N_s^3N_ta_s^3a_t}\left[\left. \frac{\partial m_f}{\partial \xi}\right|_{a_s}\langle{\rm Tr}
\left[\frac{1}{M}\right]\rangle\right. \nn\\
&&\left.+\left. \frac{\partial \nu_t}{\partial \xi}\right|_{a_s}\langle{\rm Tr}\left[\frac{\slash{D}_t}{M}\right]\rangle\right. 
+\left.\left.\frac{\partial \xi_o^{-1}}{\partial \xi}\right|_{a_s}\langle{\rm Tr}\left[\frac{\slash{D}_s}{M}\right]\rangle\right]\nn\\
p(T)&=&\frac{\varepsilon(T)}{3} + \frac{a_s}{3N_s^3N_ta_s^3a_t}\frac{1}{N_c}\times\nn\\
&&\left[\left(\frac{1}{\xi_o}\left.\frac{\partial \beta}{\partial a_s}\right|_{\xi}+\beta\left.\frac{\partial \xi_o^{-1}}
{\partial a_s}\right|_{\xi}\right)\langle \sum_{x, s>s\prime} {\rm Re} {\rm Tr}\left[ P_{ss\prime}(x)\right]\rangle\right. \nonumber\\
&&+\left.\left(\xi_o\left.\frac{\partial \beta}{\partial a_s}\right|_{\xi}+\beta\left.\frac{\partial\xi_o}{\partial a_s}\right|_{\xi}\right)\langle\sum_{x, s} {\rm Re} {\rm Tr}\left[P_{st}(x)\right]\rangle\right] \nonumber\\
&&+\frac{a_s N_f}{12N_s^3N_ta_s^3a_t}\left[\left. \frac{\partial m_f}{\partial a_s}\right|_{\xi}\langle{\rm Tr}
\left[\frac{1}{M}\right]\rangle\right.\nn\\
&&\left. +\left. \frac{\partial \nu_t}{\partial a_s}\right|_{\xi}\langle{\rm Tr}\left[\frac{\slash{D}_t}{M}\right]\rangle\right. 
+\left.\left.\frac{\partial \xi_o^{-1}}{\partial a_s}\right|_{\xi}\langle{\rm Tr}\left[\frac{\slash{D}_s}{M}\right]\rangle\right].\nn
\end{eqnarray}
}
\normalsize
\vspace{-.9cm}
\section{SIMULATIONS}
In our simulations we use the R-algorithm\cite{R-alg} implemented for two flavors of dynamical fermions.

The zero-temperature runs from Table 1 are used to calculate the derivatives of the bare parameters with respect to $\xi$ and $a_s$ 
(the Karsch coefficients) involved in the EOS.

The finite temperature runs in Table 2 and 3 represent two separate sweeps through the transition region for each of
 which the fixed bare parameter scheme described above is applied. Figure 1 shows the variation of $\langle \overline{\psi}\psi\rangle$ through the 
phase transition as we gradually change the temperature by varying only $N_t$ for $\xi =4.0(1)$ and $4.8(3)$.       

\begin{table}[ht]
\begin{small}
\begin{tabular}{llrlll}
\hline 
{\em run}& {\em volume} & {\em traj.}& $\beta$ & $\xi_0$ & $m_f$ \\ \hline

1 & $16^3$x32 &  5800   & 5.425   & 1.5    &      0.025 \\ 

2 & $16^2$x24x32 & 5100 &  5.425 &     1.5      &  0.025\\

3 & $16^2$x24x64 & 1300 &  5.695  &  2.5   &   0.025\\ 

4 & $16^2$x24x64& 1400 &  5.725 &   3.44  &   0.025 \\ 

5 & $16^2$x24x64 & 3400 & 5.6 &   3.75   & 0.025 \\ 

6 & $16^2$x24x64 & 3200  & 5.3 & 3.0 & 0.008 \\  

7 & $16^2$x24x64 & 3000  & 5.29 & 3.4 &  0.0065 \\

8 &$16^2$x24x64 &  4300   &5.286&      3.427&  0.00394\\  \hline

\end{tabular}
\caption{\small{Parameters of zero-temperature calculations. All runs have dynamical $\nu_t=1.0$ except run 3 which has
 $\nu_t=1.2$. All runs except 7 and 8 have valence $\nu_t$'s of 0.8, 1.0 and 1.2.\vspace{0.15cm}}}

\begin{tabular}{llrlll} \hline 
{\em run}& {\em volume} & {\em traj.}& $\beta$ & $\xi_0$ & $m_f$ \\ \hline
1 & $16^3$x24 & 8000  & 5.3 & 3.0 &  0.008 \\ 

2 & $16^3$x20 & 9800  & 5.3 & 3.0 & 0.008 \\  

3 & $16^3$x16 & 21600     &5.3 & 3.0 &0.008\\

4 & $16^3$x12 & 9100  & 5.3 & 3.0 & 0.008 \\

5 & $16^3$x8 & 5500  & 5.3 & 3.0 &  0.008 \\

6 & $16^3$x4 & 25900  & 5.3 & 3.0 &  0.008\\  \hline

\end{tabular}
\caption{\small{Parameters of finite temperature calculations with $\xi=4.0(1)$. All runs have dynamical $\nu_t=1.0$.\vspace{0.15cm}}}
\begin{tabular}{llrlll}\hline
{\em run}& {\em volume} & {\em traj.}& $\beta$ & $\xi_0$ & $m_f$ \\ \hline

1 & $16^3$x24 & 2200  & 5.29 & 3.4 &  0.0065 \\

2 & $16^3$x20 & 3300  & 5.29 & 3.4 &  0.0065 \\

3 & $16^3$x16 & 10300  & 5.29 & 3.4 &  0.0065 \\

4 & $16^3$x12 & 8800  & 5.29 & 3.4 &  0.0065 \\

5 & $16^3$x8 & 3600  &5.29 & 3.4 &  0.0065 \\  \hline
\end{tabular}
\caption{\small {Parameters of finite temperature calculations with $\xi=4.8(3)$. All runs have dynamical $\nu_t=1.0$.}}
\vspace{-0.6cm}
\end{small}
\end{table}

\vspace{-0.5cm}
\section{KARSCH COEFFICIENTS}

We consider the bare parameters $\xi_o$, $\beta$, $m_f$ and $\nu_t$ functions of the physical observables $\xi$, $a_s$, $R_t=(m_\pi^2/m_\rho^2)^{\rm temporal}$ and $R_{st}=(m_\pi^2/m_\rho^2)^{\rm spatial}/(m_\pi^2/m_\rho^2)^{\rm temporal}$ and expand those functions in Taylor series around the values
of the bare parameters of a selected zero-temperature run. We fit the zero-temperature data to the linear 
parts of the Taylor expansions, where the fitting parameters then include the Karsch coefficients. 
The $\chi^2$ minimization in our case is implemented with the addition that we have to start with a guess for the standard deviations 
on the bare parameters and iteratively improve this guess, yielding increasingly accurate
     values for the Karsch coefficients.
Table 4 and 5 show the numerical results for the Karsch coefficients for expansion around the bare parameters of zero-temperature runs 6 and 7.
\vspace{-0.5cm}
\begin{table}[ht]
\begin{small}
\begin{tabular}{rrrr|l}\hline
$\partial \xi_o$/&$\partial \beta$/&$\partial m_f$/&$\partial \nu_t$/&\\\hline
 0.61(6)& 9.6(6.2)&  2.0(1.1)&  0.3(2.0)& $\partial \xi$ \\
-0.017(7)& -1.5(1.1) & 0.5(2)&  0.0(3)& $\partial a_s$\\
-0.0062(4)& 0.18(5)&  0.068(4) & 0.00(2)&$\partial R_t$\\
0.04(5)& -5.6(4.8) & -0.7(8)&  1.1(1.4)&$\partial R_{st}$\\  \hline
\end{tabular}
\caption{\small{Karsch coefficients from expansion around run 6, Table 1. $\chi^2_i/D =$(1.7, 1.0 , 1.8, 0.7).\vspace{0.3cm}}}
\begin{tabular}{rrrr|l}\hline
$\partial \xi_o$/&$\partial \beta$/&$\partial m_f$/&$\partial \nu_t$/& \\\hline
0.59(6)& 9.2(6.2)&  2.2(1.0) & 0.7(1.8)&$\partial \xi$\\
-0.015(6)& -1.0(4) & 0.58(8)&  0.02(8)&$\partial a_s$\\
-0.0050(8)& 0.11(7) & 0.05(1) & 0.02(2)&$\partial R_t$\\
0.06(3) &-5.0(4.1) & -0.8(7)&  0.8(1.2)&$\partial R_{st}$\\  \hline
\end{tabular}
\caption{\small{Karsch coefficients from expansion around run 7, Table 1. $\chi^2_i/D =$(1.4, 0.7 , 2.3, 0.9).}}
\end{small}
\end{table}
\vspace{-1cm}
\section{EOS RESULTS}
Figure 2 shows the final result for the EOS. The errors on the pressure are significantly bigger than the errors on the energy due to the
large errors on those of the Karsch coefficients which are derivatives with respect to $a_s$. The comparison with the free lattice theory (squares) gives an
    explanation of the prominent drop off of $\varepsilon$ and $p$ in the
    high temperature sector --- simply a consequence of the 
    lattice high momentum cut-off.
Our results are consistent with previous isotropic results\cite{milc}.
\section{CONCLUSIONS}
We have studied the QCD thermodynamics 
using staggered fermions on anisotropic lattices.
The fixed bare parameter scheme allows us to explore the 
temperature dependence of energy and pressure with
fixed physics scales. While this approach naturally reduces finite lattice spacing errors
associated with $a_t$, the fixed lattice cut-off becomes important at increasing
temperature. Including improvements  
to the spatial parts of the staggered fermion action would be a natural step to
reduce the lattice artifacts for high temperatures.

\vspace{-.6cm}
\begin{figure}[hb]
\epsfxsize=\hsize
\epsfbox{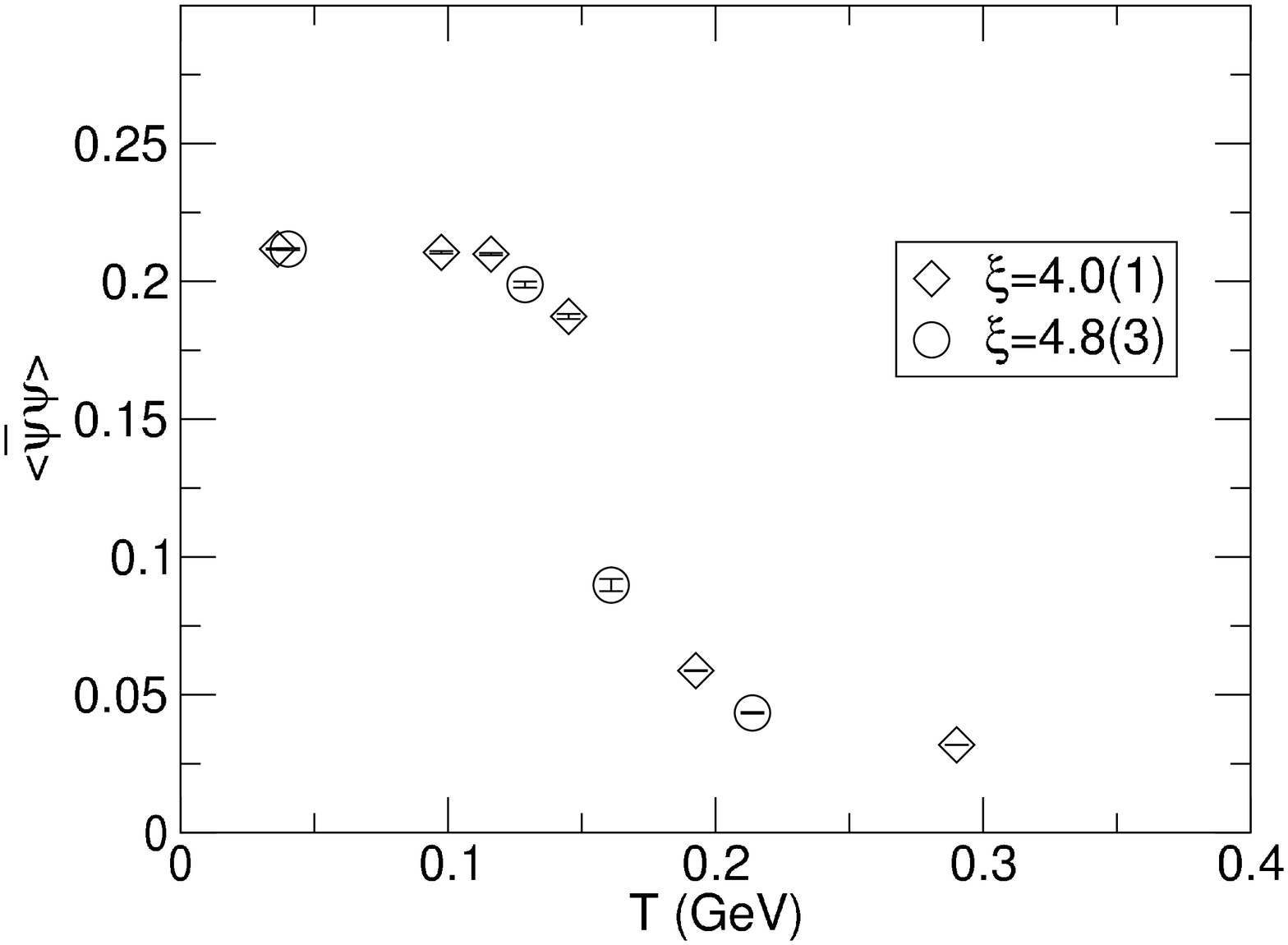}
\vspace{-1.2cm}
\caption{The temperature dependence of $\langle \overline{\psi}\psi\rangle$ in the region of $T_c$.}
\vspace{.2cm}
\epsfxsize=\hsize
\epsfbox{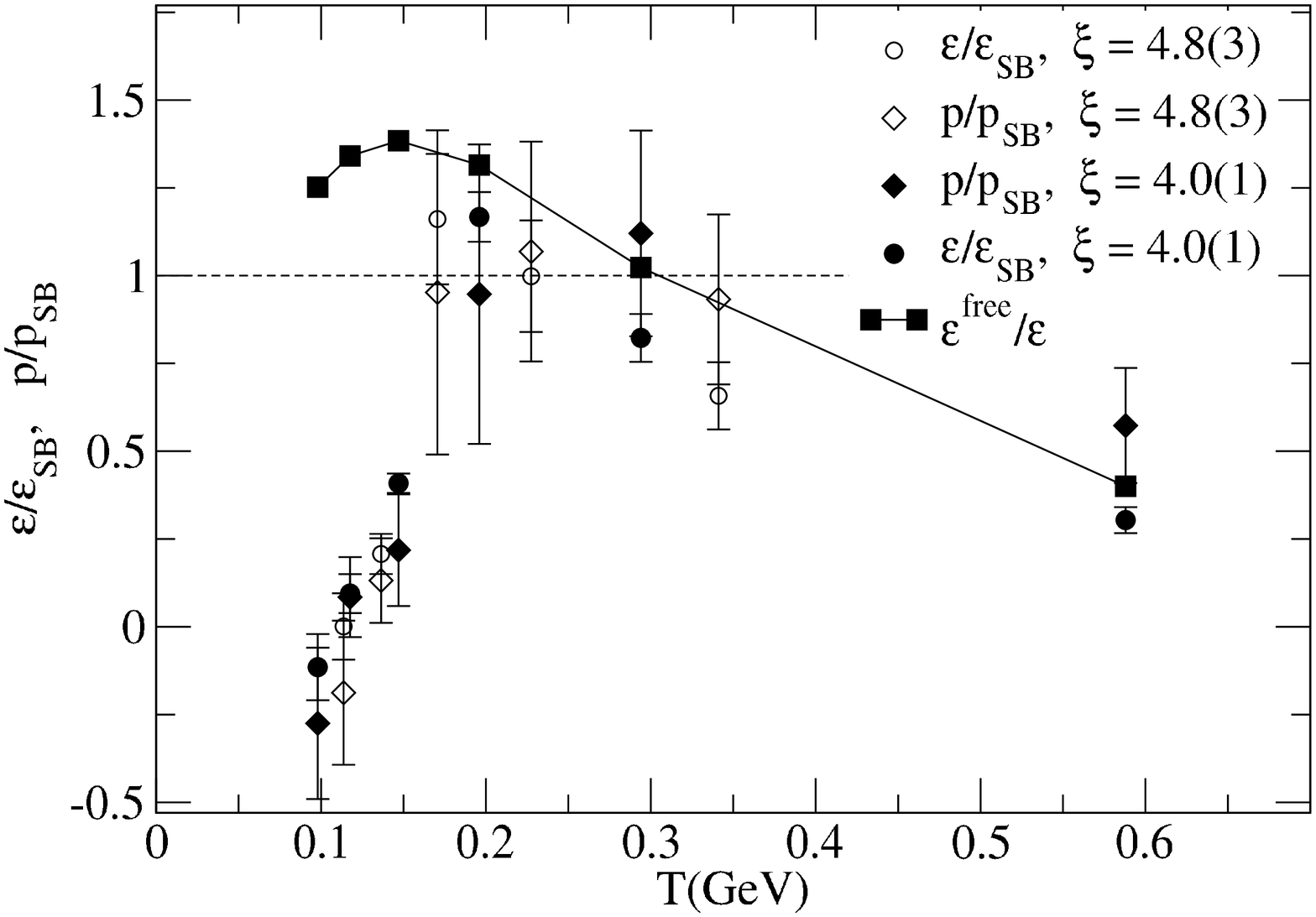}
\vspace{-1.2cm}
\caption{Energy and pressure in units of the Stefan-Boltzmann limit and a comparison with the free lattice theory (squares).}
\end{figure}
\end{document}